\documentstyle[prl,preprint,aps,epsf]{revtex}
\tighten
\begin{document}
\draft
\title{
Size-Dependent Surface Plasmon Dynamics in 
Metal Nanoparticles
}
\author{T. V. Shahbazyan and I. E. Perakis} 
\address{Department of Physics and Astronomy, 
Vanderbilt University, Box 1807-B,  
Nashville, TN 37235}
\author{J.-Y. Bigot}
\address{Institut de Physique et Chimie des Mat\'{e}riaux de Strasbourg, 
Groupe d'Optique Nonlin\'{e}aire Unit\'{e} Mixte 380046 CNRS-ULP-ECPM,
23 rue du Loess, 67037 Strasbourg Cedex, France}
\maketitle

\begin{abstract}
We study the effect of Coulomb correlations on the ultrafast optical
dynamics of small metal particles. We demonstrate that a
surface-induced dynamical screening of the electron-electron interactions  
leads to quasiparticle scattering with collective surface
excitations. In noble-metal nanoparticles, 
it results in an interband {\em resonant scattering} of $d$-holes with
surface plasmons. We show that this size-dependent many-body effect
manifests itself in the differential absorption dynamics for
frequencies close to the surface plasmon resonance. In particular, our
self-consistent calculations reveal a strong frequency dependence of
the relaxation, in agreement with recent femtosecond pump-probe
experiments.   
\end{abstract}


The absorption of light by metal nanoparticles is dominated  by the
surface plasmon  (SP) resonance. In small particles, the strong
three-dimensional confinement changes both the static and dynamic optical
properties. For example, the scattering of single-particle excitations
at the surface leads to SP damping and to a resonance linewidth inversely
proportional to the nanoparticle size\cite{kreibig}. Another 
static effect is the size-dependent enhancement of the non-linear
optical susceptibilities\cite{hac86,yan94}. 
At the same time, the effect of confinement on dynamical properties is
much less understood.\cite{kreibig,cluster}. One can  mention, for instance, 
the role of electron-electron interactions in the process of cluster
fragmentation, the role surface lattice modes in providing additional
channels for intra-molecular energy relaxation, and the influence of
the electron and nuclear motion on the superparamagnetic properties of
clusters. Ultrafast optical spectroscopy allows one to address the 
dynamical aspects by probing the time evolution of the excited states
with a resolution shorter than the dephasing or energy relaxation times.

Recently, electron relaxation in nanoparticles attracted much
interest\cite{tok94,rob95,big95,ahm96,per97,nis97}.   
Ultrafast pump-probe experiments \cite{big95,per97} performed on
noble-metal particles reported a time--dependent spectral
broadening of the SP resonance, originating from the pump--induced 
heating of the electron gas. The {\em inelastic} carrier  scattering
was found to be essential for understanding the differential transmission 
lineshape\cite{big95}. 
Similar to metal films\cite{fan92,sun94,gro95}, the dynamics
revealed an initial non-equilibrium stage during which the {\em e-e}
scattering leads to thermalization of the electron distribution,
followed by a relaxation to the lattice. 
However, certain aspects of the optical dynamics in nanoparticles 
were observed to differ  significantly from films. 
In particular, experimental studies of small Cu
nanoparticles\cite{big95} revealed considerably faster relaxation
above and below the SP resonance. These  observations suggests that,
in strongly confined systems,  collective surface excitations
should play an important role in the electron dynamics.

In this paper we address the role of collective surface excitations 
in the electron relaxation in metal nanoparticles. We demonstrate
that the {\em e-e} interactions {\em inside} small particles are strongly
modified due to a surface-induced dynamical screening. This 
leads to a possibility of quasiparticle scattering
accompanied by the emission of SP's. We show, in particular, that in
noble-metal particles, this novel size-dependent many-body mechanism
results in a resonant scattering of the $d$-hole into the conduction
band. The latter effect manifests itself in a strong frequency
dependence of the optical dynamics close to the SP resonance, and can
be  observed experimentally with ultrafast pump-probe spectroscopy.

We consider spherical metal particles, embedded in a medium with
dielectric constant $\epsilon_m$, with radii $R$ sufficiently
small, so that only dipole surface modes can be optically  excited.
The optical properties of such a colloid are determined by the
dielectric function,   
$\epsilon_m+3p\epsilon_m(\epsilon-\epsilon_m)/(\epsilon+2\epsilon_m)$,
where $\epsilon(\omega)=\epsilon'(\omega)+i\epsilon''(\omega)$ is 
that of a single metal particle and $p\ll 1$ is the volume fraction
occupied by the nanoparticles. 
For noble metals, $\epsilon(\omega)$ includes the (complex) interband
dielectric function, $\epsilon_d(\omega)$, associated with the
$d$-electrons. The absorption  coefficient is then given by\cite{kreibig} 

\begin{equation}\label{absor}
\alpha(\omega)= -9p{\omega\over c}\epsilon_m^{3/2}
\mbox{Im} {1\over \epsilon_{sp}(\omega)},
\end{equation}
with

\begin{equation}\label{epseff}
\epsilon_{sp}(\omega)=
\epsilon_d(\omega)-\omega_p^2/\omega(\omega+i\gamma_s)+2\epsilon_m.
\end{equation}
Here  $\gamma_s\sim v_F/R$ characterizes the SP damping by intraband
{\em e-h} excitations due to the surface
scattering, and $\omega_p$ is the plasmon frequency of the conduction
(s-p) electrons. The SP frequency, $\omega_{sp}$, is determined by  
$\epsilon'_{sp}(\omega_{sp})=0$

Equation (\ref{absor}) describes the linear absorption
by large metal clusters. When the electron gas is in the
quasiequilibrium with the lattice, the temperature dependence of
$\alpha(\omega)$ also determines the time evolution of the differential
absorption measured by pump-probe spectroscopy. This is the case for
time delays between the pump and the probe optical pulses longer than 
$\sim 1$ ps \cite{fan92,sun94,gro95}. 
Since for electron temperature $T$ much smaller than the Fermi
energy $E_F$, the intraband contribution to $\epsilon_{sp}(\omega)$
is nearly $T$-independent, the SP dynamics comes mainly from the
$T$-dependence of the {\em interband} dielectric function
$\epsilon_d(\omega)$. The latter is typically approximated by its
bulk expression calculated for non-interacting
electrons\cite{kreibig}. In such model, the SP dynamics is essentially
size-independent. 

Here we point out  that in small particles,
there is a size-dependent contribution to $\epsilon_d(\omega)$ that 
arises from an interband $d$-hole  scattering mechanism, absent in
bulk metals. This many-body effect  originates from the fact that the
collective surface excitations strongly  modify the dynamically screened
{\em e-e} interactions inside nanoparticles, as we demonstrate below.  

The screened Coulomb potential at point ${\bf r}$ arizing from 
an electron at point ${\bf r}'$ inside a spherical particle,
$U_{\omega}({\bf r},{\bf r}')$,  
is determined by the equation\cite{mahan}

\begin{eqnarray}\label{dyson}
U_{\omega}({\bf r},{\bf r}')=U({\bf r}-{\bf r}')+
%&&
\int d{\bf r}_1 d{\bf r}_2U({\bf r}-{\bf r}_1)
%\nonumber\\&&\times
\Pi_{\omega}({\bf r}_1,{\bf r}_2)U_{\omega}({\bf r}_2,{\bf r}'),
\end{eqnarray}
where $U({\bf r}-{\bf r}')$ is the unscreened  Coulomb potential and
$\Pi_{\omega}({\bf r}_1,{\bf r}_2)$ is the polarization
operator. The latter includes contributions from the conduction and
$d$-band, as well as from the surrounding medium,
$\Pi_{\omega}=\Pi_{\omega}^c+\Pi_{\omega}^d+\Pi_{\omega}^m$. 
For simplicity,  we assume here step density profiles,
$n_{i}({\bf r})=\bar{n}_{i}\theta(R-r)$, $i=c,d$, where 
$\bar{n}_{c}$ ($\bar{n}_{d}$) is the average density of the conduction
(d-) electrons. For the high frequencies of interest, $\Pi_{\omega}^c$
can be expanded in $1/\omega$, yielding to lowest order\cite{lus74} 
$\Pi_{\omega}^c({\bf r},{\bf r}_1)=
-(1/m\omega^2)\nabla [n_c({\bf r})\nabla \delta({\bf r}-{\bf r}_1)]$.
The $d$-band contribution to the rhs of Eq.\ (\ref{dyson}) can be
found from the relation
 
\begin{equation}\label{delp}
-e^2\int d{\bf r}_2\Pi_{\omega}^d({\bf r},{\bf r}_1)
U_{\omega}({\bf r}_1,{\bf r}')
=\nabla {\bf P}_{\omega}^d({\bf r},{\bf r}'),
\end{equation}
where ${\bf P}_{\omega}^d({\bf r},{\bf r}')=-\chi_d(\omega)
\nabla U_{\omega}({\bf r},{\bf r}')$ is the $d$-band  polarization
vector (${\bf P}_{\omega}^d=0$ outside the nanoparticle) and
$\chi_d=(\epsilon_d-1)/4\pi$ is the interband  
susceptibility. $\Pi_{\omega}^m$ can be obtained from a similar relation
(with ${\bf P}_{\omega}^m=0$ inside the nanoparticle).
After some algebra, 
Eq.~(\ref{dyson}) takes the form

%\begin{eqnarray}\label{self1}
\begin{equation}\label{self1}
\left(\epsilon_d  - {\omega_p^2\over \omega^2}\right)
%&&
U_{\omega}({\bf r},{\bf r}')=
U({\bf r}-{\bf r}')+
\left({\epsilon_d-\epsilon_m\over 4\pi}-{e^2\bar{n}_c\over m\omega^2}\right)
%\nonumber\\ \times
\int d{\bf r}_1 
%&&
\nabla_1{1\over |{\bf r}-{\bf r}_1|}\nabla_1\theta(R-r_1)
U_{\omega}({\bf r}_1,{\bf r}'),
%\end{eqnarray}
\end{equation}
with $\omega_p^2=4\pi e^2\bar{n}_c/m$.
We solved Eq.~(\ref{self1}) by expanding $U_{\omega}$ in
spherical harmonics, $Y_{LM}(\hat{\bf r})$. The solution reads\cite{shaper}

\begin{eqnarray}\label{screen}
U_{\omega}({\bf r},{\bf r}')
%&&
={U({\bf r}-{\bf r}')\over\epsilon(\omega)}
+ {e^2\over R}\sum_{LM}{4\pi\over 2L+1}
\left({rr'\over R^2}\right)^L 
%\nonumber\\ && \times
Y_{LM}(\hat{\bf r})Y^{\ast}_{LM}(\hat{\bf r}')
\left[{1\over \epsilon_L(\omega)}
-{1\over\epsilon(\omega)}\right],
\end{eqnarray}
where $\epsilon(\omega)=\epsilon_d(\omega)-\omega_p^2/\omega^2$. Here
$\epsilon_L(\omega)=\epsilon'_L(\omega)+i\epsilon''_L(\omega)$ 
is an effective dielectric function, whose zeros, 
$\epsilon'_L(\omega)={L\over 2L+1} 
[\epsilon'_d(\omega)-\omega_p^2/\omega^2]
+{L+1\over 2L+1}\epsilon_m=0$
determine the frequencies of the multipole collective surface
excitations\cite{kreibig}. Note that $\epsilon''_L(\omega)$
contains a correction due to the  damping of collective excitations by
electron-hole pairs, which can be obtained by adding 
${\rm Im}\Pi_{\omega}^c$ to the rhs of Eq.~(\ref{self1}).

Equation (\ref{screen}) represents a generalization of the plasmon
pole approximation for  spherical particles. 
The two terms in the rhs describe two distinct
contributions. First is the usual  bulk-like screening of the Coulomb
potential. The second, {\em size-dependent}  contribution descibes a new 
effective {\em e-e} interaction induced by the {\em surface}: the
potential of an electron  inside the nanoparticle excites collective
surface modes, which in turn act as image charges that interact with
the second electron. For frequencies close to that of the SP, 
it is sufficient to retain only the resonant dipole term, $L=1$. The
corresponding surface-induced {\em e-e} interaction potential
is then 

\begin{eqnarray}\label{spscreen}
U^{sp}_{\omega}({\bf r},{\bf r}')=
3{e^2\over R}{{\bf r}\cdot {\bf r}'\over R^2}
{1\over \epsilon_{sp}(\omega)},
\end{eqnarray}
where $\epsilon_{sp}(\omega)=3\epsilon_1(\omega)$ is the same as in 
Eq.~(\ref{epseff}). Thus, for sufficiently small $R$, the dynamically 
screened Coulomb potential in the nanoparticle is dominated by
the SP pole. This leads to the possibility of quasiparticle scattering
accompanied by the emission of a SP. 

In particular, consider the SP-mediated scattering of the $d$-hole
into the conduction band. This process is described by the Matsubara
self-energy due to the potential (\ref{spscreen})\cite{mahan}

\begin{eqnarray}\label{mself}
\Sigma_{\alpha}^{d}(i\omega)=
-3{e^2\over R^3}\sum_{\alpha'}
{1\over \beta}\sum_{i\omega'}
{|{\bf d}_{\alpha\alpha'}|^2\over \epsilon_{sp}(i\omega')}
G_{\alpha'}^{c}(i\omega'+i\omega).
\end{eqnarray}
Here 
${\bf d}_{\alpha\alpha'}=\langle c,\alpha |{\bf r}|d,\alpha'\rangle=
\langle c,\alpha |{\bf p}|d,\alpha'\rangle/im(E^c_{\alpha}-E^d_{\alpha'})$ 
(we set $\hbar=1$), where $|c,\alpha'\rangle$ ($|d,\alpha'\rangle$) and 
$E^{c}_{\alpha}$ ($E^{d}_{\alpha}$) are the conduction ($d$-) band
eigenstates and eigenenergies, respectively,  and $G_{\alpha}^{c}$ is
the Green function of noninteracting conduction electrons. 
For a quasicontinuous spectrum, the matrix element can be approximated
by a bulk-like expression, $\langle c,\alpha |{\bf p}|d,\alpha'\rangle
=\delta_{\alpha\alpha'}\langle c|{\bf p}|d\rangle
\equiv \delta_{\alpha\alpha'}\mu$, the corrections due 
to the surface scattering being $\sim R^{-1}$. After the 
frequency summation, Im$\Sigma_{\alpha}^{d}$ takes the form

\begin{eqnarray}\label{imself1}
{\rm Im}\Sigma_{\alpha}^{d}(E)
=-{9e^2 \mu^2\over m^2(E^{cd}_{\alpha})^2R^3}
\mbox{Im}{N(E^c_{\alpha}-E)+f(E^c_{\alpha})
\over\epsilon_{sp}(E^{c}_{\alpha}-E)},
\end{eqnarray} 
where $E^{cd}_{\alpha}=E^{c}_{\alpha}-E^{d}_{\alpha}$, and 
$f(E)$ and $N(E)$ are the Fermi and Bose distributions, respectively.

We see from Eq.\ (\ref{imself1}) that the scattering rate of a 
$d$-hole with energy $E^{d}_{\alpha}$, 
$\gamma_{h}(E^{d}_{\alpha})=\mbox{Im}\Sigma_{\alpha}^{d}(E^{d}_{\alpha})$,
depends strongly on the nanoparticle size. Importantly, 
it exhibits a {\em peak} as the interband separation
$E^{cd}_{\alpha}$ approaches the SP frequency 
$\omega_{sp}$. The reason for such a sharp energy dependence 
is that the surface potential (\ref{spscreen}) induces only vertical
(dipole) transitions, thus restricting the number of allowed final
states. 

Let us turn to the absorption spectrum. Consider an optically excited
{\em e-h} pair with excitation energy $\omega$ close to
the onset of interband transitions, $\Delta$. As discussed above, the
$d$-hole can subsequently scatter into the conduction band by emitting
a SP. Eq.~(\ref{imself1}) indicates that for
$\omega\sim\omega_{sp}\sim\Delta$, 
this process will be resonantly enhanced. In particular, this is the
case in Cu and Au, where $\omega_{sp}$ and $\Delta$ are close to each
other\cite{big95,per97}.  At the same time, the electron
scatters in the conduction band  via the usual
two-quasiparticle process. For $\omega\sim\Delta$, the electron energy
is close to $E_F$, and the scattering rate is 
$\gamma_{e}\sim 10^{-2}$ eV \cite{petek}. Using the bulk value of $\mu$,
$2\mu^2/m\sim 1$ eV near the L-point\cite{eir62}, we find that 
$\gamma_{h}$ exceeds $\gamma_{e}$ for nanoparticles 
smaller than $\sim 5$\ nm.  

We now calculate the  effect of the SP-assisted interband
scattering on the absorption coefficient $\alpha$.
For non-interacting electrons, the interband susceptibility, 
$\chi_d(i\omega)=\tilde{\chi}_d(i\omega)+\tilde{\chi}_d(-i\omega)$,
has the standard form
\begin{equation}\label{susc}
\tilde{\chi}_d(i\omega)=
-\sum_{\alpha}{e^2\mu^2\over m^2(E^{cd}_{\alpha})^2}
{1\over \beta}\sum_{i\omega'}
G_{\alpha}^{d}(i\omega')
G_{\alpha}^{c}(i\omega'+i\omega).
\end{equation}
Since the $d$-band is occupied for all energies, it is sufficient to
include only the effect of $d$-hole scattering 
into $\tilde{\chi}_d(i\omega)$ \cite{zan81}. Substituting  
$G_{\alpha}^{d}(i\omega')= 
[i\omega'-E^{d}_{\alpha}+E_F-\Sigma_{\alpha}^{d}(i\omega')]^{-1}$,
with $\Sigma_{\alpha}^{d}(i\omega)$ given by Eq.~(\ref{mself}),
we obtain after the frequency summation

\begin{equation}\label{interband}
\tilde{\chi}_d(\omega)=
{e^2 \mu^2\over m^2}\int {dE^c\, g(E^c)\over (E^{cd})^2}
{f(E^c)-1\over \omega-E^{cd}+i\gamma_h(\omega,E^c)},
\end{equation}
where $g(E^c)$ is the density of states in the conduction band, and
we assumed a dispersionless $d$-band with energy $E^d$. Here
 $\gamma_h(\omega,E^c)={\rm Im}\Sigma^{d}(E^c-\omega)$ is the
scattering rate of a $d$-hole with energy $E^c-\omega$,

\begin{equation}\label{gamhole}
\gamma_h(\omega,E^c)
=-{9e^2 \mu^2\over m^2(E^{cd})^2R^3}
f(E^c)\mbox{Im}{1\over\epsilon_{sp}(\omega)}.
\end{equation}
The comparison of  Eqs.\ (\ref{gamhole}) and (\ref{absor}) shows that
the scattering rate {\em depends on the frequency} in the same way as the
absorption coefficient: both $\gamma_h(\omega,E^c)$ and
$\alpha(\omega)$ exhibit a peak at $\omega=\omega_{sp}$. This implies
that the $d$-hole experiences {\em resonant scattering} into the
conduction band as the optical field frequency approaches the SP
frequency. We thus arrive at a self-consistent problem defined 
by Eqs.\ (\ref{absor}), (\ref{epseff}), (\ref{interband}), and
(\ref{gamhole}).
 
The main role of the SP-assisted $d$-hole scattering is
to change the absorption lineshape in the vicinity of the SP
resonance. The SP width is determined by
$\epsilon_d''(\omega)=4\pi\tilde{\chi}_d(\omega)$, 
which now, according to Eqs. (\ref{gamhole}) and (\ref{interband}),
acquires a sharp frequency dependence for $\omega \sim \omega_{sp}$.
It is also important that the effect of $\gamma_h$ on
$\epsilon_d''(\omega)$ increases with temperature.
Indeed, the Fermi function in the rhs of Eq.\ (\ref{gamhole}) implies that 
$\gamma_h$ is small unless $E^c-E_F \lesssim k_B T$. Since the main
contribution to $\mbox{Im}\chi_d^{res}(\omega)$ comes from energies 
$E^c-E_F \sim \omega-\Delta$, the scattering becomes efficient for
temperatures $k_B T\gtrsim \omega_{sp}-\Delta$. The combined effect of
the above $\omega$ and $T$ dependence manifests itself strongly in the
differential absorption dynamics, as illustrated below.

In Fig.~1 we show the results of our self-consistent numerical 
calculations of the differential transmission dynamics. We use the
parameters of the experiment\cite{big95}, performed on $\sim 5$ nm
Cunanoparticles, with the SP frequency,  
$\omega_{sp}\simeq 2.22$ eV, slightly above the onset of the
interband transitions, $\Delta\simeq 2.18$ eV.
Fig.~1(a) shows the spectra at various time delays, calculated as 
the difference between $\alpha(\omega)$  at $T=300$ K and  $T(t)$, 
the pump-induced hot electron temperature. The latter was obtained
from the two-temperature model with bulk Cu parameters\cite{eas86} and
initial temperature $T_0=800$ K. The $d$-hole scattering leads to
a steeper $\omega$-dependence of the differential transmission for
$\omega>\omega_{sp}$; its effect, however, is best seen in the time
evolution as described below. We also included the effect of the intraband
{\em e-e} scattering; for electron energy close to $E_F$, this  can be
achieved  by adding the {\em e-e} scattering rate\cite{pines} 
$\gamma_e(E^c)\propto [1-f(E^c)][(E^c-E_F)^2+(\pi k_B T)^2]$ to
$\gamma_h$ in Eq.~(\ref{interband}). The difference in 
$\gamma_e(E^c)$ for $E^c$ smaller and larger than $E_F$ leads to an
asymmetric differential transmission lineshape, in agreement
with experiment\cite{big95}. Note that in small nanoparticles,
$\gamma_e$ is expected to exceed its bulk value\cite{siv94}, which
could account for the more symmetric lineshape observed in 40 nm
nanoparticles\cite{per97}.  

In Fig.~1(b) we show the calculated time evolution of the differential
transmission. The relaxation is slowest at $\omega=\omega_{sp}$, which
can be attributed to the robustness of the SP mode; 
the corresponding relaxation time is 3.5 ps. More importantly, for
short time delays, the relaxation is significantly faster below 
{\em and} above $\omega_{sp}$, with a characteristic time of 1.4
ps. These results are consistent with the experimental data  shown
here for comparison in Fig.~1(c) (the experiment also revealed an
initial non-equilibrium  regime not addressed here). In contrast, in
the absence of $d$-hole scattering  [$\gamma_h=0$ in
Eq.~(\ref{interband})], the frequency dependence would be 
smooth {\em above} the resonance [see Fig.~1(d)]. Such a smooth 
dependence originates from the lineshape of the absorption peak:
since $\omega_{sp}\sim\Delta$, the  absorption is larger for
$\omega>\omega_{sp}$ due to the interband transitions, and therefore  
the {\em relative} change in the absorption is smaller. This, however,
changes with $d$-hole scattering turned on ($\gamma_h\neq 0$). To 
understand this difference, one should note that away from
$\omega_{sp}$, the relaxation characterizes the temperature-induced
change in the SP {\em width}. 
The latter acquires an additional frequency dependence near
the resonance, originating from that of $\gamma_h(\omega)$.  
This effect is enhanced above the resonance  due to the similar
frequency dependence of $\gamma_h$ and $\alpha$. Furtheremore, 
the efficiency of this mechanism increases with $T$, as discussed
above. This results in  a strong frequency dependence  of the
relaxation (for $\omega>\omega_{sp}$) for shorter time delays, which
correspond to higher T [see Fig.~1(b)].  

In conclusion, we have shown that surface-induced
electron-electron interactions in small metal particles lead to
quasiparticle scattering mediated by  collective surface
excitations. In noble-metal particles, this size-dependent mechanism
results in a surface-plasmon-assisted resonant scattering 
of $d$-holes into the conduction band. The latter effect can be
observed with ultrafast  pump-probe spectroscopy. Our work points out
the increasing role of many-body correlations with decreasing size,
important for  understanding the transition from 
boundary-constrained nanoscale materials to molecular clusters.

This work was supported by NSF CAREER award ECS-9703453, and, in
part, by ONR Grant N00014-96-1-1042  and by Hitachi Ltd.

\begin{figure}
\caption{
(a) Calculated differential transmission spectra for positive time
delays with initial hot electron temperature $T_0=800$ K,
$\epsilon_m=2.25$, and $\gamma_s=0.1$ eV.
(b) Temporal evolution of the differential transmission 
for frequencies equal, above, and below the SP resonance. 
(c) Experimental temporal evolution of the pump-probe
signal \protect\cite{big95}. 
(d) Calculated  differential transmission in the absence of interband 
$d$-hole scattering.
}
\end{figure}

\clearpage

\begin{figure}[b]
\epsfxsize=6.0in
\epsffile{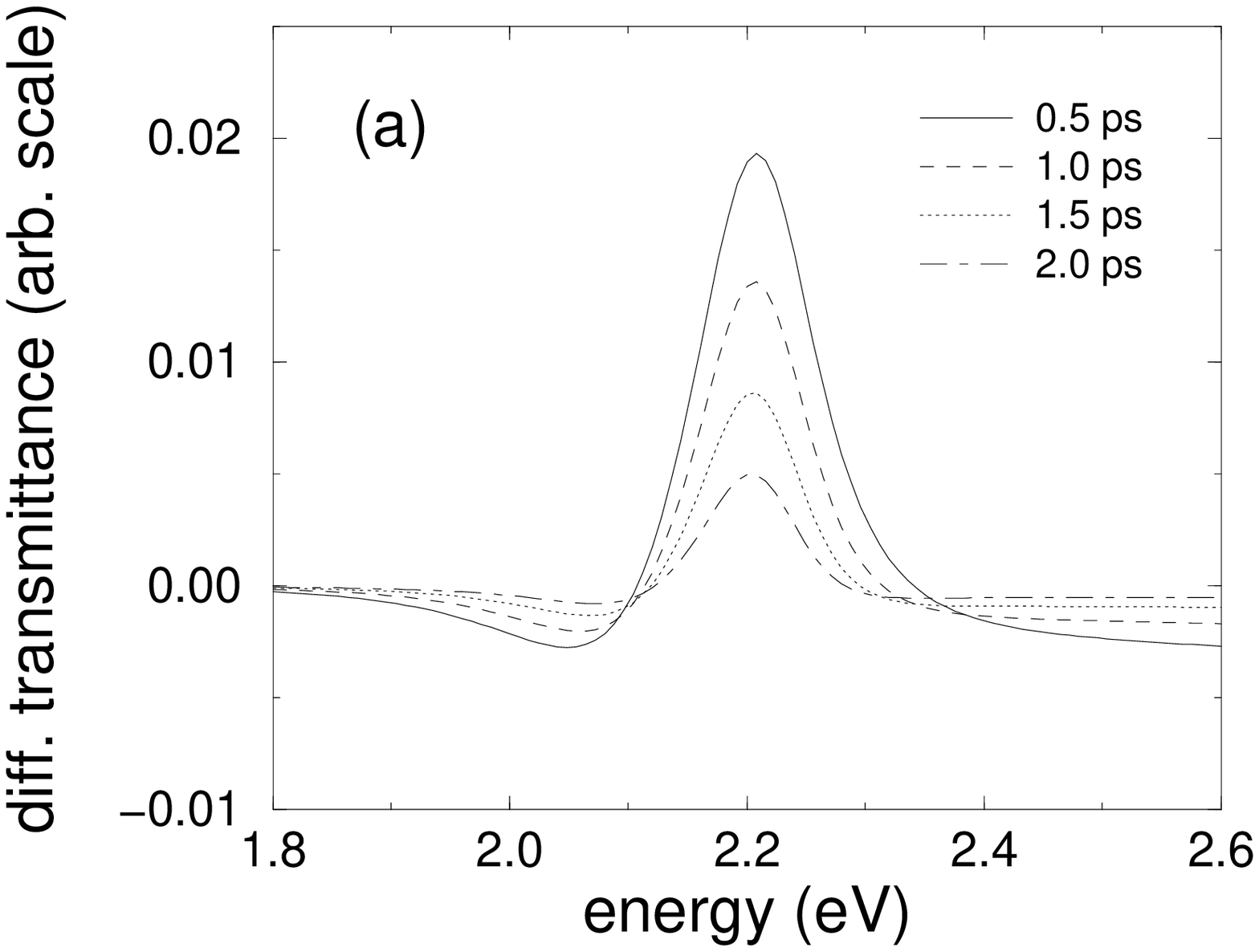}
\epsfxsize=6.0in
\epsffile{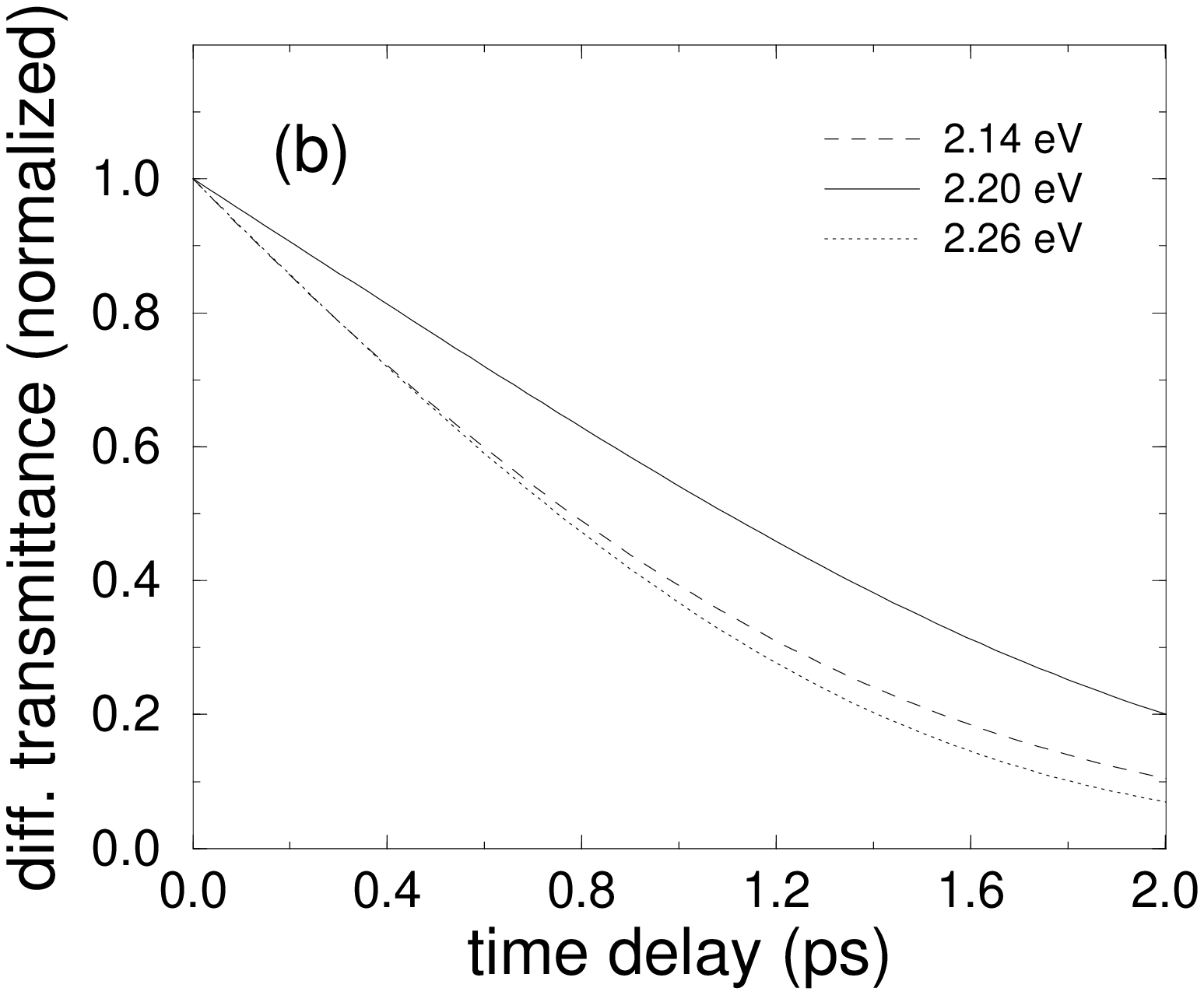}
\epsfxsize=6.0in
\epsffile{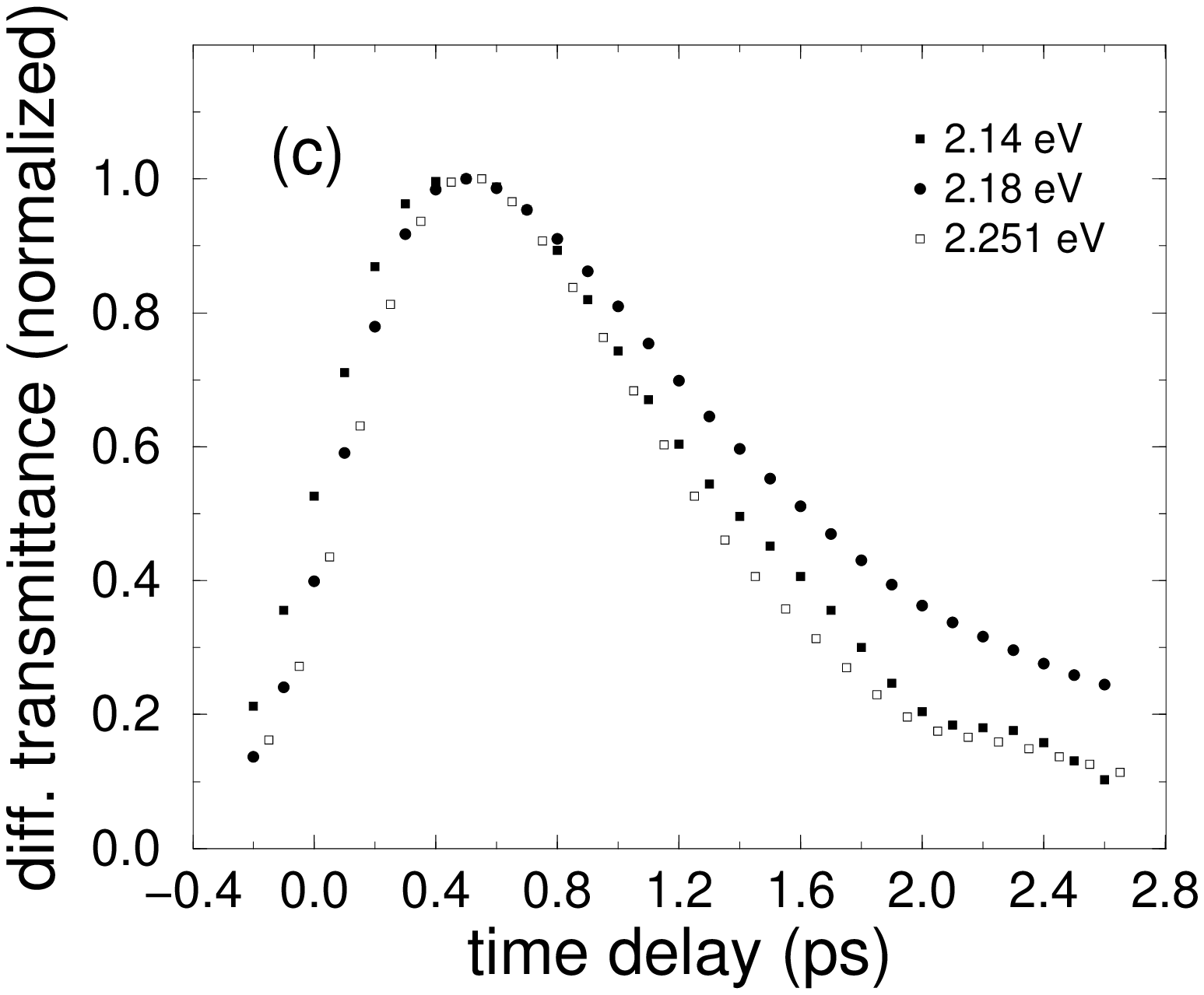}
\epsfxsize=6.0in
\epsffile{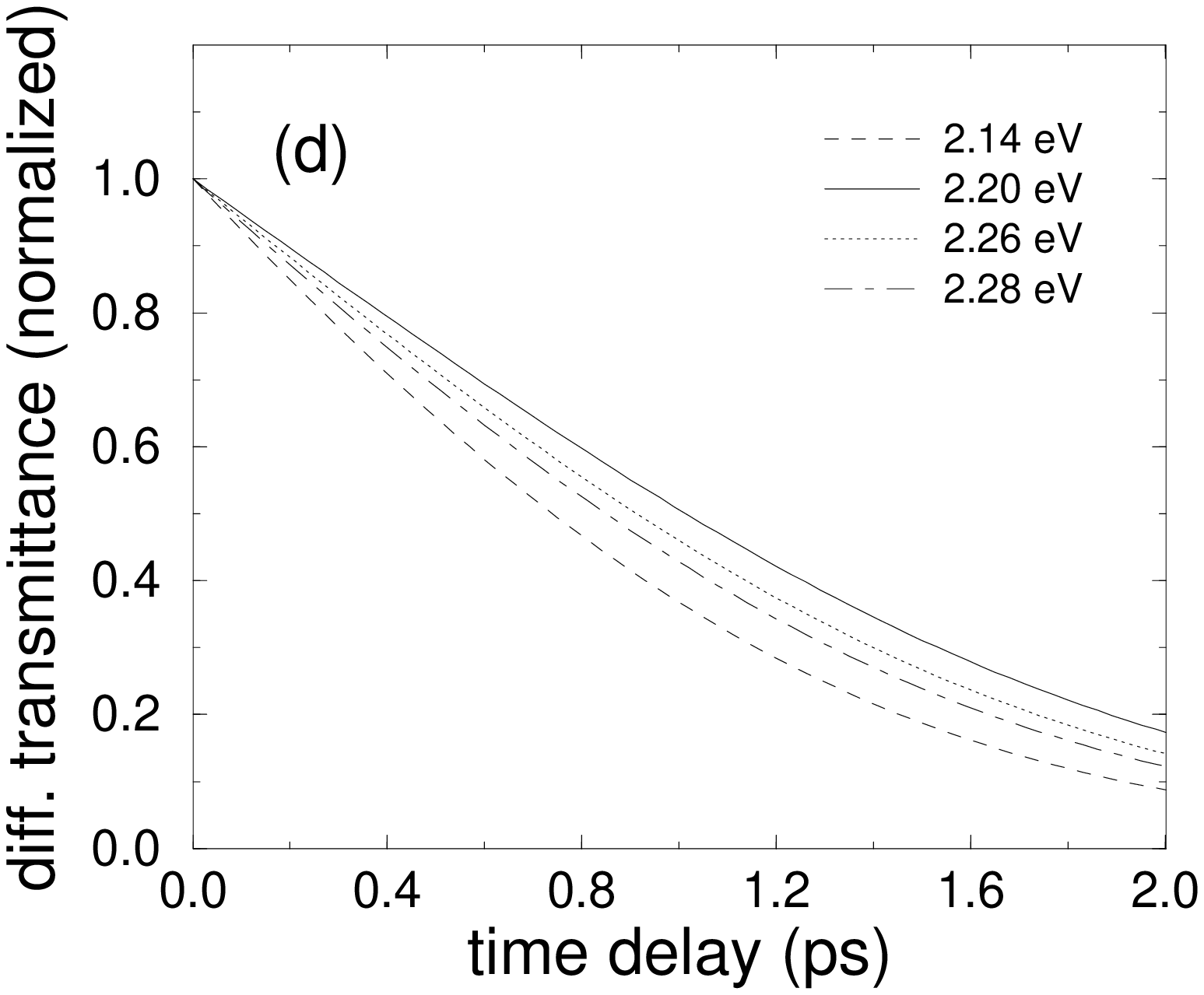}
\label{fig.1}
\end{figure}

\end{document}